\documentclass[10pt,conference,letterpaper]{IEEEtran}
\usepackage{caption}
\usepackage{amsfonts,amssymb}
\usepackage{amsmath}
\usepackage{bm}
\usepackage{graphicx}
\usepackage{epstopdf}
\usepackage{subfigure}
\usepackage{ragged2e}
\usepackage{cite}
\usepackage{verbatim}
\usepackage{tablefootnote}

\begin{document}

\title{Fairness Optimization of RSMA for Uplink Communication based on Intelligent Reflecting Surface}

\author{\IEEEauthorblockA{Shanshan Zhang, and  Wen Chen}
	
	\IEEEauthorblockA{Department of Electronic Engineering, Shanghai Jiao Tong University, China\\
		Email: $\lbrace$ shansz,wenchen $\rbrace$@sjtu.edu.cn}}

\maketitle

\begin{abstract}
In this paper, we propose a rate-splitting multiple access (RSMA) scheme for uplink wireless communication systems with intelligent reflecting surface (IRS) aided. In the considered model, IRS is adopted to overcome power attenuation caused by path loss. We construct a max-min fairness optimization problem to obtain the resource allocation, including the receive beamforming at the base station (BS) and phase-shift beamforming at IRS.  We also introduce a successive group decoding (SGD) algorithm at the receiver, which trades off the fairness and complexity of decoding. In the simulation, the results show that the proposed scheme has superiority in improving the fairness of uplink communication.

\end{abstract}


\IEEEpeerreviewmaketitle

\section{Introduction}

\IEEEPARstart{}{} 
With the development of the sixth-generation mobile communications (6G) system, wireless networks need to support massive connectivity and provide services with higher throughput, ultra-reliability, and heterogeneous quality of service (QoS). Therefore, wireless systems must make more efficient use of wireless resources and manage interference more rigorously. To address these issues, rate splitting multiple access (RSMA) has been proposed as a design of physical layer (PHY) and multiple access technique \cite{RN450}. It splits the signal into two streams at the transmitter to manage the interference. Different from non-orthogonal multiple access (NOMA), which fully decodes the interference from other devices, RSMA partially decodes the interference and partially treats them as the noise at the receiver. Therefore, RSMA provides a new paradigm for massive connectivity, which bridges the two extremes of fully decoding interference and fully treating interference as noise \cite{9831440,RN476}. It has been recognized as a promising scheme for non-orthogonal transmission, interference management, and massive access strategies in 6G \cite{9832618}. 

Recently, several works have studied the RSMA scheme. \cite{RN408} studies sum-rate maximization for different communication systems. Some work \cite{RN449, RN339} focus on researching the performance of RSMA with imperfect channel state information at the transmitter (CSIT) in the network. 
\cite{RN474} jointly optimize the parameters of IRS and RSMA to improve energy efficiency and spectral efficiency. For RSMA-based robust and secrecy communication systems, \cite{RN419,9963667} studied the sum-rate maximization and fairness design. But most of the current work is concerned with downlink communications. The uplink RSMA is proposed in \cite{RN365} which proved that  RSMA can achieve the capacity region of the Gaussian multiple access channel (MAC) without time sharing among devices.  There are several works that investigate uplink communications. The uplink RSMA schemes are applied to improve outage performance \cite{RN407} and fairness \cite{RN368} in a two-device MAC. 
\cite{RN426}  focused on joint optimization of power allocation to the uplink devices and beamforming design to maximize the sum rate for the uplink RSMA system. 
 Based on existing work, the performance of RSMA is strongly dependent on the successive interference cancellation (SIC) \cite{RN365, RN338}. But in the uplink system with massive connectivity, RSMA faces the challenges of complexity issues and SIC processing delay. Therefore, implementing RSMA in uplink wireless networks also faces several problems such as decoding schemes and resource management for message transmission. 

In order to reduce the complexity at the receiver and the time delay of the signal processing, we apply successive group decoding (SGD) in the uplink RSMA system. SGD is introduced in \cite{RN438}, which is an extension of the conventional SIC. In SGD, a subset of devices can be jointly decoded instead of just one at each decoding stage. Therefore, SGD can reduce the complexity of decoding at the base station (BS) and take advantage of RSMA to improve fairness. Another promising technique for 6G is intelligent reflecting surface (IRS), which is applied to improve the network coverage and to resolve blockage issues in wireless communications \cite{RN475,9950724, RN473}. \cite{RN426} studied the downlink RSMA scheme with IRS-aided to achieve better rate performance and enhanced coverage capability for multiple devices. Therefore, to meet the demand QoS of future communication, SGD and IRS can be utilized to allocate resources and solve this fairness issue.
 
In this paper, we propose an IRS-aided uplink RSMA framework that adopts the SGD scheme at the receiver. The IRS assists the direct transmission from devices to the BS and improves spectral and energy efficiencies. With SGD, the RSMA can achieve any point in the capacity region of a Gaussian multiple-access channel. We formulate a max-min fairness optimization problem to jointly optimize the design of grouping order, receive beamforming at the BS, and phase-shift beamforming at IRS. To solve the optimization problem, we adopt an alternating optimization (AO) algorithm to iteratively optimize the receive beamforming and phase-shift beamforming. Then we give a greedy grouping algorithm with low complexity to design the group decoding order to achieve fairness. Numerical results show the proposed IRS-aided RSMA transmission framework based on SGD improves the worst-case rate among devices compared with other schemes without SGD. Therefore, the proposed  RSMA framework improves fairness and is more powerful than the existing transmission schemes.

\section{System Descriptions}
In this section, we first structure the IRS-aided uplink RSMA system. Then we introduce the SGD algorithm at the receiver. 

\subsection{System Model}
\label{SM}
We consider an uplink RSMA system, which consists of $K$ single-antenna devices, a BS equipped with $M$ antennas, and an IRS composed of $N$ elements. In the RSMA, the $K$ original messages are split into $2K$ sub-messages. Denote $x_{k,i}$ as the $i$th sub-message of device $k$, where $i=1,2$. Accordingly, power constraints are assigned to these sub-inputs to satisfy the original constraints. 
$p_{k,i}$ denotes the power allocation for the $i$th sub-message for device $k$, where $i=1,2$. Each device $k$ has a maximum transmit power limit $P_{\text{max}}$, i.e.,$\sum\limits_{i=1}^2 p_{k,i}\leq P_{\text{max}}$. 
 The received signal $\bm y\in\mathbb{C}^{M\times 1}$ at the BS is 
\begin{eqnarray}	
\bm y= \sum\limits_{k=1}^K\left(\mathbf H_{rb}\mathbf\Theta\bm h_{sr,k}+\bm h_{d,k}\right)\left(\sqrt{p_{k,1}}x_{k,1}+\sqrt{p_{k,2}}x_{k,2}\right)+\bm w\nonumber,
\end{eqnarray}
where $\mathbf H_{rb}\in \mathbb{C}^{M\times N}$, $\bm h_{sr,k}\in \mathbb{C}^{N\times 1}$, and $\bm h_{d,k}\in \mathbb{C}^{M\times 1}$ are the channels from IRS to the BS, device $k$ to the IRS, and device $k$ to the BS, respectively. $\mathbf\Theta=\text{diag} [e^{j\theta_1},\ldots,e^{j\theta_N}]$ is the phase-shift matrix, where $\theta_n\in(-\pi,\pi]$ is the phase shift induced by the $n$th element of the IRS. $\bm w\sim\mathcal{CN}(\bm 0,\sigma^2\mathbf I)$ is the additive white Gaussian noise (AWGN).
The massive MIMO system adopts a block-fading model where channels follow independent quasi-static flat-fading in each block of coherence time. According to \cite{RN115}, the channel matrix $\mathbf H_{rb}$ is given by 
 \begin{eqnarray}
\mathbf H_{rb} = \sum\limits_{p=1}^{N_{rb}}\beta_{p}^{rb}\bm a_{B}(\theta_{B,p}^{rb})\bm a_{R}^H(\theta_{R,p}^{rb})e^{-j2\pi\tau_{p}^{rb} \frac{B_s}{2}},
\label{eq:2}
\end{eqnarray}
where $B_s$ represents the two-sided bandwidth, and $N_{rb}$ denotes the number of multi-path components (MPCs). $\beta_{p}^{rb}$ and $\tau_{p}^{rb}$ are the complex path gain and the path delay of the $p$th MPC, respectively. The array steering and response vectors are given by
\begin{eqnarray}
 \begin{aligned}
\bm a_{B}(\theta_{B,p}^{rb})= [1, e^{-j2\pi\theta_{B,p}^{rb}},\ldots,e^{-j2\pi(M-1)\theta_{B,p}^{rb}}]^T,\\
\bm a_{R}(\theta_{R,p})= [1, e^{-j2\pi\theta_{R,p}^{rb}},\ldots,e^{-j2\pi(N-1)\theta_{R,p}^{rb}}]^T.
\label{eq:3}
 \end{aligned}
\end{eqnarray}
$\theta_{\cdot,p}^{rb}$ is related to the physical angle $\phi_{\cdot,p}^{rb}\in [-\pi/2,\pi/2]$ as $\theta_{\cdot,p}^{rb}=d\sin(\phi_{\cdot,p}^{rb})/\lambda$, where $\lambda$ is the wavelength of propagation, $d$ is the antenna spacing with $d=\lambda$. Similarly, $\bm h_{sr,k}$ and $\bm h_{d,k}$ are given by 
 \begin{eqnarray}
 \begin{aligned}
\bm h_{sr,k} &= \sum\limits_{p=1}^{N_{sr,k}}\beta_{p,k}^{sr}\bm a_{R}(\theta_{R,p,k}^{sr})e^{-j2\pi\tau_{p,k}^{sr}\frac{B_s}{2}},\\
\bm h_{d,k} &= \sum\limits_{p=1}^{N_{d,k}}\beta_{p,k}^{d}\bm a_{B}(\theta_{B,p,k}^{d})e^{-j2\pi\tau_{p,k}^{d}\frac{B_s}{2}},
\label{eq:4}
\end{aligned}
\end{eqnarray}
where $N_{sr,k}$ and $N_{d,k}$ denote the number of MPCs for the channel from device $k$ to the IRS, and device $k$ to the BS. $\beta_{p,k}^{sr}$, $\beta_{p,k}^{d}$, $\tau_{p,k}^{sr}$, and $\tau_{p,k}^{d}$ are the complex path gain and the path delay of the $p$th MPC for the channel from device $k$ to the IRS, and device $k$ to the BS, respectively. 

Then we have
\begin{eqnarray}
\begin{aligned}	
\bm h_k&=\mathbf H_{rb}\mathbf\Theta\bm h_{sr,k}+\bm h_{d,k}
=\mathbf H_k\bm v,
\end{aligned}
\end{eqnarray}
where $\mathbf H_k=[\mathbf H_{rb}\text{diag}(\bm h_{sr,k});\bm h_{d,k}]\in\mathbb C^{M\times(N+1)}$ and $\bm v=[\text{diag}(\mathbf\Theta);1]=[e^{j\theta_1},\ldots,e^{j\theta_N},1]^T$.
Finally, the received signal is written as 
\begin{eqnarray}
 \begin{aligned}	
\bm y
=\sum\limits_{k=1}^K\mathbf H_k\bm v\left(\sqrt{p_{k,1}}x_{k,1}+\sqrt{p_{k,2}}x_{k,2}\right)+\bm w.
 \end{aligned}
\end{eqnarray} 

\subsection{Successive Group Decoding (SGD)}
\label{SGD}
In the SGD, a subset of sub-messages is decoded with treating the transmissions of the undecoded sub-messages as interference at each stage. Define the sub-message set $\mathcal Q\triangleq\{(k,i)\}_{k=1,i=1}^{K,I}$.
Assume that the devices' sub-messages are divided into $L$ groups, i.e., $\mathcal Q_l\triangleq\{(k,i)|(k,i)\in\mathcal Q\}, l=1,\ldots, L$. There are $\mathcal Q_1\cup\cdots\cup\mathcal Q_L=\mathcal Q$ and $\mathcal Q_1\cap\cdots\cap\mathcal Q_L=\emptyset$. The decoding order at the BS is $\mathcal Q_1,\ldots,\mathcal Q_L$. If $(k,i)\in\mathcal Q_l$, 
the BS will employ linear beamforming on the received signal $\bm y$ for decoding sub-message ${x}_{k,i}$,
\begin{eqnarray}
 \begin{aligned}	
\hat{x}_{k,i}=&\bm g_{k,i}^H \bm y\\=&\bm g_{k,i}^H\mathbf H_k\bm v \sqrt{p_{k,i}} x_{k,i}+\bm g_{k,i}^H\bm w\\+&\bm g_{k,i}^H\sum\limits_{(n,j)\in\mathcal Q_l\cup\ldots\cup Q_L,(n,j)\neq (k,i)}\mathbf H_n\bm v\sqrt{p_{n,j}}x_{n,j} ,
\label{eq:x}
 \end{aligned}
\end{eqnarray} 
where $\bm g_{k,i}\in\mathbb C^{M\times 1}$ denotes the beamforming vector for the $i$th sub-message of device $k$. The SGD operates as follows.
\begin{itemize}
	\item a) Initialize with inputs: $l=1,\mathbf H_1,\ldots,\mathbf H_K,$ and $\mathcal Q_1,\ldots,\mathcal Q_L$.
	\item b) For $(k,i)\in \mathcal Q_l$, estimate $x_{k,j}$ according to (\ref{eq:x}).
	\item c) Update $\bm y = \bm y - \sum\limits_{(k,i)\in\mathcal Q_l}\mathbf H_k\bm v\left(\sqrt{p_{k,i}}\hat x_{k,i}\right)$ and $l=l+1$.
	\item d) If $l=L+1$, stop, otherwise go to step b).
\end{itemize}
Therefore, the rate of $(k,i)\in \mathcal Q_l$ is expressed as (\ref{eq:xhat}).
\begin{figure*}[!t]
	\vspace*{-8pt}
	\normalsize
\begin{eqnarray}	
r_{k,i}=\log_2\left(1+\frac{p_{k,i}\|\bm g_{k,i}^H\mathbf H_k\bm v\|_2^2}{\sum\limits_{(n,j)\in\mathcal Q_l\cup\cdots\cup\mathcal Q_L,(n,j)\neq(k,i)}p_{n,j}\|\bm g_{k,i}^H\mathbf H_n\bm v\|_2^2+\|\bm g_{k,i}\|_2^2\sigma^2}\right).
\label{eq:xhat}
\end{eqnarray}
	\hrulefill
	\vspace*{-8pt}
\end{figure*}

\section{Resource Allocation for Fairness}
In this section, we focus on fair rate adaptation for the IRS-aided uplink network. Specifically, we formulate the fair rate adaptation as a max-min problem under the joint consideration of decoding order and beamforming design (including receive beamforming at the BS and phase-shift beamforming at IRS).

\subsection{Problem Formulation}
To maximize the minimum rate among all devices, we formulate the joint design of receive beamforming at the BS, phase-shift beamforming at IRS, grouping order of decoding. The max-min problem is as follows:
\begin{eqnarray}
\begin{aligned}
P0:\quad \max_{\mathcal Q_l,\bm v ,\bm g_{k,i}}& \quad \min_{(k,i)\in\mathcal Q} r_{k,i}\\
s.t. & \quad \sum\limits_{i=1}^2 \|\bm g_{k,i}\|_2^2\leq P_{\text{max}}, \forall k, \\&\quad |[\bm v]_n| =1, n=1,\ldots,N,[\bm v]_{N+1}=1,
\end{aligned}
\end{eqnarray}
where $P_{\text{max}}$ is the maximum power limit. 
To facilitate the solution design, we define $\mathbf G_{k,i}=\bm g_{k,i}\bm g_{k,i}^H, \mathbf V=\bm v\bm v^H$, where $\mathbf G_{k,i}\succeq\mathbf 0$, $\text{rank}(\mathbf G_{k,i})\leq 1,\forall (k,i)\in\mathcal Q$, $\mathbf V\succeq\mathbf 0$, and $\text{rank}(\mathbf V)\leq 1$. Then the rate can be rewritten as (\ref{eq:r}). 
\begin{figure*}[!t]
	\vspace*{-8pt}
	\normalsize
	\begin{eqnarray}	
	r_{k,i}=\log_2\left(1+\frac{p_{k,i}\text{tr}(\mathbf H_k\mathbf V\mathbf H_k^H\mathbf G_{k,i})}{\sum\limits_{(n,j)\in\mathcal Q_l\cup\cdots\cup\mathcal Q_L,(n,j)\neq(k,i)}p_{n,j}\text{tr}(\mathbf H_n\mathbf V\mathbf H_n^H\mathbf G_{k,i} )+\text{tr}(\mathbf G_{k,i})\sigma^2}\right).
	\label{eq:r}
	\end{eqnarray}
	\vspace*{-8pt}
\end{figure*}
Give the definition $u_{k,i}$ and $d_{k,i}$ in (\ref{eq:u}) and (\ref{eq:d}), respectively.
\begin{figure*}[!t]
	\vspace*{-8pt}
	\normalsize
\begin{align}
u_{k,i}&\triangleq\log_2\left(\sum\limits_{(n,j)\in\mathcal Q_l\cup\cdots\cup\mathcal Q_L}p_{n,j}\text{tr}(\mathbf H_n\mathbf V\mathbf H_n^H\mathbf G_{k,i} )+\text{tr}(\mathbf G_{k,i})\sigma^2\right) \label{eq:u}\\
d_{k,i}&\triangleq\log_2\left(\sum\limits_{(n,j)\in\mathcal Q_l\cup\cdots\cup\mathcal Q_L,(n,j)\neq(k,i)}p_{n,j}\text{tr}(\mathbf H_n\mathbf V\mathbf H_n^H\mathbf G_{k,i} )+\text{tr}(\mathbf G_{k,i})\sigma^2\right).\label{eq:d}
\end{align} 
\hrulefill
\vspace*{-8pt}
\end{figure*}
Then we have $r_{k,i}=u_{k,i}-d_{k,i}$. Finally, we introduce auxiliary variables $r$ and equivalently convert (P0) into the following form,
\begin{eqnarray}
\begin{aligned}
P1:\quad \max_{\mathcal Q_l,\mathbf V ,\mathbf G_{k,i},r}& \quad r\\
s.t. \quad(C1)&\quad \sum\limits_{i=1}^2 \text{tr}(\mathbf G_{k,i})\leq P_{\text{max}}, \forall k, \\(C2)&\quad \mathbf G_{k,i}\succeq\mathbf 0, \text{rank}(\mathbf G_{k,i})\leq 1,\forall (k,i)\in\mathcal Q, \\(C3)&\quad [\mathbf V]_{nn}=1, n=1,\ldots,N+1,\\(C4)&\quad \mathbf V\succeq\mathbf 0, \text{rank}(\mathbf V)\leq 1,\\(C5)&\quad u_{k,i}-d_{k,i}\ge r, \forall (k,i)\in\mathcal Q,\nonumber
\end{aligned}
\end{eqnarray}
where (C1) is the power constraint of receive beamforming. (C2) and (C4) impose semidefinite and nonnegativity constraints of $\mathbf G_{k,i}$ and $\mathbf V$, respectively. (C3) ensures the unit-modulus constraints on the phase shifts. It is evident that (P1) is an intractable non-convex problem due to the coupled optimization variables in the objective function and the non-convex unit-modulus constraints in (C3). Therefore, this non-convex problem is hard to solve directly.

To solve this optimization problem, we adopt the AO algorithm, which is widely used and empirically efficient for driving the non-convex problem with coupled optimization variables. Specifically, we decouple (P1) into three sub-problems, i.e., receive beamforming optimization, phase-shift beamforming optimization, and decoding order optimization. Then we alternately optimize the three sub-problems until convergence is achieved.

\subsection{Optimizing Receive Beamforming }
\label{opt_G}
We aim to optimize receive beamforming for given phase-shift beamforming and decoding order. Therefore, for given $\mathbf V$ and $\mathcal Q_l$,  the subproblem of (P1) is
\begin{eqnarray}
\begin{aligned}
P2:\quad &\max_{\mathbf G_{k,i},r} \quad r\qquad
s.t. \quad&(C1),(C2),(C5).
\end{aligned}\label{eq:p1.1}
\end{eqnarray}
Note that the functions $u_{k,i}$ and $d_{k,i}$ are concave for $\mathbf G_{k,i}$ and the concavity of $d_{k,i}$ makes the objective function non-convex. The iterative successive convex approximation (SCA) is used to address this problem. Specifically, we use SCA to linearly approximate the $d_{k,i}$ as follows
\begin{eqnarray}
\begin{aligned}	
&d_{k,i}(\mathbf G_{k,i})\leq d_{k,i}(\mathbf G_{k,i}^r)\\&\qquad\qquad+\text{tr}\left(\left(\nabla_{\mathbf G_{k,i}}d_{k,i}\left(\mathbf G_{k,i}^r\right)\right)^T\left(\mathbf G_{k,i}-\mathbf G_{k,i}^r\right)\right)\\&\qquad\qquad\triangleq d_{k,i}^r(\mathbf G_{k,i}),
\end{aligned}\label{eq:appr1}
\end{eqnarray}
where 
$\nabla_{\mathbf G_{k,i}}d_{k,i}(\mathbf G_{k,i}^r)=\frac{\left(\sum\limits_{(n,j)\in\mathcal Q_l\cup\cdots\cup\mathcal Q_L,(n,j)\neq(k,i)}p_{n,j}\mathbf H_n\mathbf V\mathbf H_n^H+\mathbf I\sigma^2\right)^T}{\left(\sum\limits_{(n,j)\in\mathcal Q_l\cup\cdots\cup\mathcal Q_L,(n,j)\neq(k,i)}p_{n,j}\text{tr}(\mathbf H_n\mathbf V\mathbf H_n^H\mathbf G_{k,i}^r )+\text{tr}(\mathbf G_{k,i}^r)\sigma^2\right)\ln2}$ and $\mathbf G_{k,i}^r$ is the local feasible point in the $r$-th iteration. Eq.(\ref{eq:appr1}) gives an upper-bounded of $d_{k,i}$ by its first-order Taylor expansion. Therefore, (C5) can be approximately transformed into
\begin{align}
\quad u_{k,i}(\mathbf G_{k,i})-d_{k,i}^r(\mathbf G_{k,i})\ge r, \forall (k,i)\in\mathcal Q.
\end{align}
However, due to the non-convex rank constraints in (C2), problem (P2) is still a non-convex problem. To address this problem, we exploit the penalty-based method \cite{RN416} to handle the rank constraint. To be specific, 
\begin{align}
\text{rank}(\mathbf G_{k,i})\leq 1\Rightarrow \text{tr}(\mathbf G_{k,i})-\|\mathbf G_{k,i}\|_2=0.
\end{align} 
Then, we incorporate the constraint $\text{tr}(\mathbf G_{k,i})-\|\mathbf G_{k,i}\|_2=0$ into the objective function (P2) by introducing a positive penalty parameter $\rho_1$, and obtain the problem (P2.1) as follows
\begin{eqnarray}
\begin{aligned}
P2.1:\max_{\mathbf G_{k,i},r}& \quad  r-\frac{1}{2\rho_1}\sum\limits_{(k,i)\in \mathcal Q}\left(\text{tr}\left(\mathbf G_{k,i}\right)-\|\mathbf G_{k,i}\|_2\right),\\
s.t.(C1)&\quad \sum\limits_{i=1}^2 \text{tr}(\mathbf G_{k,i})\leq P_{\text{max}}, \forall k,\\(\overline{C2})&\quad \mathbf G_{k,i}\succeq\mathbf 0,\forall (k,i)\in\mathcal Q.\\(\overline{C5})&\quad u_{k,i}(\mathbf G_{k,i})-d_{k,i}^r(\mathbf G_{k,i})\ge r, \forall (k,i)\in\mathcal Q.
\end{aligned}\label{eq:1.2}
\end{eqnarray}
According to Theorem 1, problem (P2.1) can obtain a rank-one solution when $\rho_1$ is sufficiently small. 

\textbf{Theorem 1:} Let $\{\mathbf G_{k,i}^s\}_{(k,i)\in\mathcal Q}$ be the optimal solution of (P2.1) with penalty parameter $\rho_s$. When $\rho_s$ is sufficiently small, i.e., $\rho_s\to 0$, then any set of limit  points $\{\bar{\mathbf G}_{k,i}\}_{(k,i)\in\mathcal Q}$ of the sequence $\{\{\mathbf G_{k,i}^s\}_{(k,i)\in\mathcal Q}\}$ is an optimal solution of problem (P2).

{\it{Proof:}} Please refer to Appendix \ref{appendix:t1}.

Note that the convexity of $\|\mathbf G_{k,i}\|_2$ makes problem (P2.1) still non-convex. Therefore, we replace $\|\mathbf G_{k,i}\|_2$
with a lower bound given by a first-order Taylor expansion of $\|\mathbf G_{k,i}\|_2$, i.e.,
\begin{align}
\|\mathbf G_{k,i}\|_2\geq \|\mathbf G_{k,i}^r\|_2+\text{tr}\left(\bm\alpha_{k,i}^r\bm(\bm\alpha_{k,i}^r)^H\left(\mathbf G_{k,i}-\mathbf G_{k,i}^r\right)\right),\nonumber
\end{align}
where $\bm\alpha_{k,i}^r$ represents the  eigenvector which corresponds to the largest eigenvalue of $\mathbf G_{k,i}^r$. Then we can approximate problem (P2.1) as
\begin{eqnarray}
\begin{aligned}
P2.2:\max_{\mathbf G_{k,i},r}& \quad  r-\frac{1}{2\rho_1}\sum\limits_{(k,i)\in \mathcal Q}(\text{tr}\left(\mathbf G_{k,i}\right)-\|\mathbf G_{k,i}^r\|_2\\&\quad-\text{tr}\left(\bm\alpha_{k,i}^r\bm(\bm\alpha_{k,i}^r)^H\left(\mathbf G_{k,i}-\mathbf G_{k,i}^r\right)\right)),\\
s.t.\quad&(C1), (\overline{C2}), (\overline{C5}).
\end{aligned}\label{eq:1.3}
\end{eqnarray}
It is observed that (P2.2) is a convex semidefinite program (SDP) that can be efficiently solved by off-the-shelf solvers such as the CVX toolbox.

\subsection{Optimizing Phase-shift Beamforming}
\label{opt_V}
In this part, we aim to optimize phase-shift beamforming for given receive beamforming and decoding order. For given $\mathbf G_{k,i}$ and $\mathcal Q_l$, (P1) is reduced to 
\begin{eqnarray}
\begin{aligned}
P3:\quad& \max_{\mathbf V,r } \quad r\qquad
s.t. \quad&(C3),(C4),(C5).
\end{aligned}\label{eq:p1.4}
\end{eqnarray}
As in Section \ref{opt_G}, we apply the SCA method to tackle
this problem. Specifically, by applying the first-order Taylor
expansion to $d_{k,i}(\mathbf V)$, we obtain
\begin{eqnarray}
\begin{aligned}	
d_{k,i}(\mathbf V)&\leq d_{k,i}(\mathbf V^r)+\text{tr}\left(\left(\nabla_{\mathbf V}d_{k,i}(\mathbf V^r)\right)^T\left(\mathbf V-\mathbf V^r\right)\right)\\&\triangleq d_{k,i}^r(\mathbf V),
\end{aligned}
\end{eqnarray}
where $\nabla_{\mathbf V}d_{k,i}(\mathbf V^r)=\\ \frac{\left(\sum\limits_{(n,j)\in\mathcal Q_l\cup\cdots\cup\mathcal Q_L,(n,j)\neq(k,i)}p_{n,j}\mathbf H_n^H\mathbf G_{k,i}\mathbf H_n\right)^T}{\left(\sum\limits_{(n,j)\in\mathcal Q_l\cup\cdots\cup\mathcal Q_L,(n,j)\neq(k,i)}p_{n,j}\text{tr}(\mathbf H_n\mathbf V^r\mathbf H_n^H\mathbf G_{k,i} )+\text{tr}(\mathbf G_{k,i})\sigma^2\right)\ln2}$ and $\mathbf V^r$ is the local feasible point in the $r$-th iteration.

The only remaining obstacle to solving problem (P3) is the non-convex rank constraint (C4). As in the case of optimizing receive beamforming, we can also exploit the penalty-based method to handle the rank constraint. To be specific, 
\begin{align}
\text{rank}(\mathbf V)\leq 1\Rightarrow \text{tr}(\mathbf V)-\|\mathbf V\|_2=0.
\end{align} 
Then, we incorporate the constraint $\text{tr}(\mathbf V)-\|\mathbf V\|_2=0$ into the objective function (P3) by introducing a positive penalty parameter $\rho_2$, which yields the following problem
\begin{eqnarray}
\begin{aligned}
P3.1:\max_{\mathbf V,r}& \quad  r -\frac{1}{2\rho_2}(\text{tr}\left(\mathbf V\right)-\|\mathbf V\|_2)\\
s.t. (C3)&\quad [\mathbf V]_{nn}=1, n=1,\ldots,N+1,\\(\overline{C4})&\quad \mathbf V\succeq\mathbf 0,\\(\overline{C5})&\quad u_{k,i}(\mathbf V)-d_{k,i}^r(\mathbf V)\ge r, \forall (k,i)\in\mathcal Q.
\end{aligned}
\end{eqnarray}
According to Theorem 1, problem (P3.1) can obtain a rank-one solution when $\rho_2$ is sufficiently small. Note that the convexity of $\|\mathbf V\|_2$ makes problem (P3.1) still non-convex. 
Denote $\bm\lambda^r$ represents the eigenvector that corresponds to the largest eigenvalue of $\mathbf V^r$. Then we can replace $\|\mathbf V\|_2$
with a lower bound given by a first-order Taylor expansion of $\|\mathbf V\|_2$, i.e.,
\begin{align}
\|\mathbf V\|_2\geq \|\mathbf V^r\|_2+\text{tr}\left(\bm\lambda^r\bm(\bm\lambda^r)^H\left(\mathbf V-\mathbf V^r\right)\right).
\end{align}
The problem (P3.1) can be approximated as
\begin{eqnarray}
\begin{aligned}
P3.2:\max_{\mathbf V,r}& \quad  r -\frac{1}{2\rho_4}(\text{tr}\left(\mathbf V\right)-\|\mathbf V^r\|_2\\&\qquad-\text{tr}\left(\bm\lambda_{max}^r(\bm\lambda_{max}^r)^H\left(\mathbf V-\mathbf V^r\right)\right))\\
s.t. \quad&(C3), (\overline{C4}), (\overline{C5}),
\end{aligned}
\end{eqnarray}
(P3.2) is a convex SDP and can be solved by existing CVX.

\subsection{Optimizing Decoding Order}
 
As we describe in Section \ref{SGD}, SGD allows multiple devices to be decoded jointly by the linear detection method in each group, and it can have a higher rate than direct decoding by linear detection methods. The decoding order is important in SGD because it will decide the rate achievable region of the system. Thus the objective of this part is to identify the group decoding orders of devices under the constraints. For given receive beamforming $\mathbf G_{k,i},\forall (k,i)\in\mathcal Q$ and phase-shift beamforming $\mathbf V$, problem (P1) can be written as 
\begin{eqnarray}
\begin{aligned}
P4:\quad &\max_{\mathcal Q_l,r} \quad r\qquad
s.t. \quad&(C5).
\end{aligned}\label{eq:p1.7}
\end{eqnarray} 
In this section, we develop an algorithm to specify the decoding order. In the proposed algorithm, the decoding order at the BS is determined in a greedy fashion. Based on Section \ref{SGD}, in the $l$-th stage of the SGD, sub-massages in group $\mathcal Q_l$ are decoded. Therefore, in the $l$-th stage, we select the combination of sub-messages that have the minimum sum rate over all combinations of undecided sub-messages, i.e., 
\begin{eqnarray}
\begin{aligned}
\mathcal Q_l=\arg\min\limits_{\mathcal V\subset\mathcal Q\setminus  \{\mathcal Q_k\},k<l,|\mathcal V|=q}\sum\limits_{(k,i)\in\mathcal V^c} r_{k,i},
\end{aligned}
\end{eqnarray}
where $\mathcal V^c=\mathcal Q\setminus  \{\mathcal Q_k,\mathcal V\},k<l$ and  $q$ is the size of $\mathcal Q_l$ \footnote{We can also give limit the size of $\mathcal Q_l$ as $|\mathcal Q_l|\le q_{\text{max}}$, but this will lead to increased complexity. To alleviate the complexity, we set the fixed number of sub-messages in each decoding group to $q$.}. The greedy grouping algorithm is summarized in Algorithm 1.
\begin{table}[ht]
		\vspace{-0.2cm}
	\normalsize
	\centering  
	\setlength{\tabcolsep}{7mm}{
		\begin{tabular}{l}  
			\hline  
			\textbf{Algorithm 1:} Greedy Grouping Algorithm \\
			\hline	
			1: Initialize: $l=1$, $\underline{\mathcal Q}=\emptyset$, $\mathcal S=\emptyset$, and $\mathcal G = \mathcal Q$.
			\\
			2: {Repeat} \\
			3: If $l<L$,\\ \qquad$\mathcal Q_l=\arg\min\limits_{\mathcal V\subset\mathcal G,|\mathcal V|=q}\sum\limits_{(k,i)\in\mathcal V^c} r_{k,i},$\\ \qquad$l=l+1$,\\
			\quad else\\ \qquad $\mathcal Q_l\gets\mathcal G$,\\
			4: $\mathcal G\gets \mathcal G\backslash \mathcal Q_l$, $\underline{\mathcal Q}\gets\{\mathcal Q_l,\underline{\mathcal Q}\}$,\\
			5: Until $\mathcal G=\emptyset$. \\
			6: {Return} $\underline{\mathcal Q}$.\\
			\hline
		\end{tabular}
	}
	\vspace{-0.3cm}
\end{table}

\subsection{Computational Complexity Analysis}
In each iteration, (P2.2) and (P3.3) optimize beamforming by the interior point method, so the computational complexities are $\mathcal O(M^{3.5})$ and $\mathcal O((N+1)^{3.5})$, respectively. (P4) is solved by Algorithm 1, whose complexity can be represented by $\mathcal O(L)$. Therefore,  the computational complexity of the proposed algorithm can be expressed as $\mathcal O(r((N+1)^{3.5}+M^{3.5}+L))$, where $r$ is the number of iterations.

\section{Simulation Results}

This section shows the simulation results of the proposed AO  algorithm. We set the antenna $M=16$, the reflecting element of IRS $N=16$, and bandwidth $B_s=1$MHz. The complex path delay and the path gain follows $\tau_{[\cdot,\cdot]}^{[\cdot]}\sim U(0,1/B_s)$ and $\beta_{[\cdot,\cdot]}^{[\cdot]}\sim \mathcal{CN}(0,1)$, respectively. The simulation results illustrate that SGD can improve the fairness of the uplink rate. 

\begin{figure}[t]
	\centering
	\begin{center}
		\includegraphics[scale=0.3]{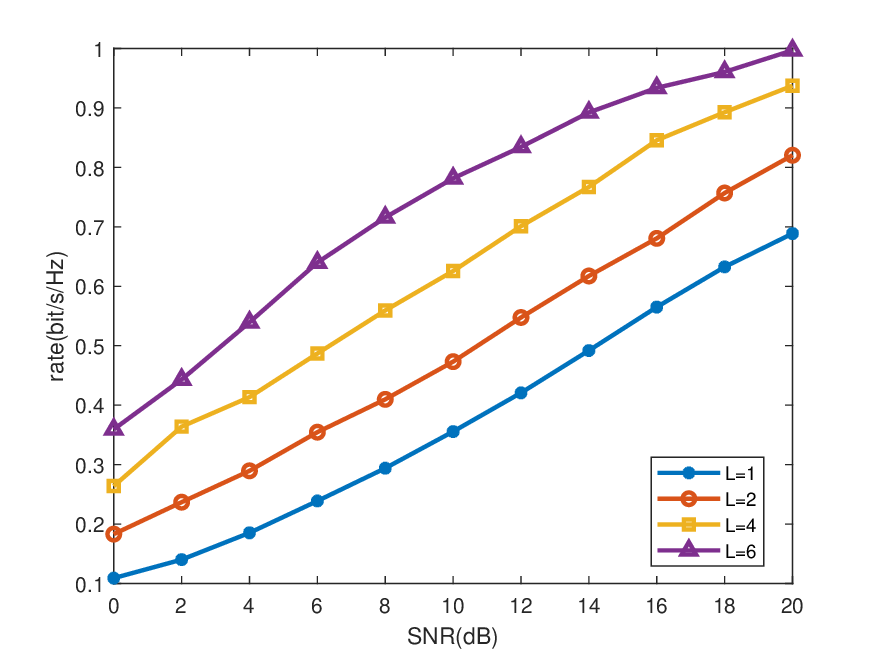}
	\end{center}
	\vspace{-0.4cm}
	\caption{The minimum rate versus SNR with different $L$. There is $K=6$ and $p_{k,i}=1/2$.}
	\label{fig1}
	\vspace{-0.5cm}
\end{figure}
\begin{figure}[t]
	\centering
	\begin{center}
		\includegraphics[scale=0.3]{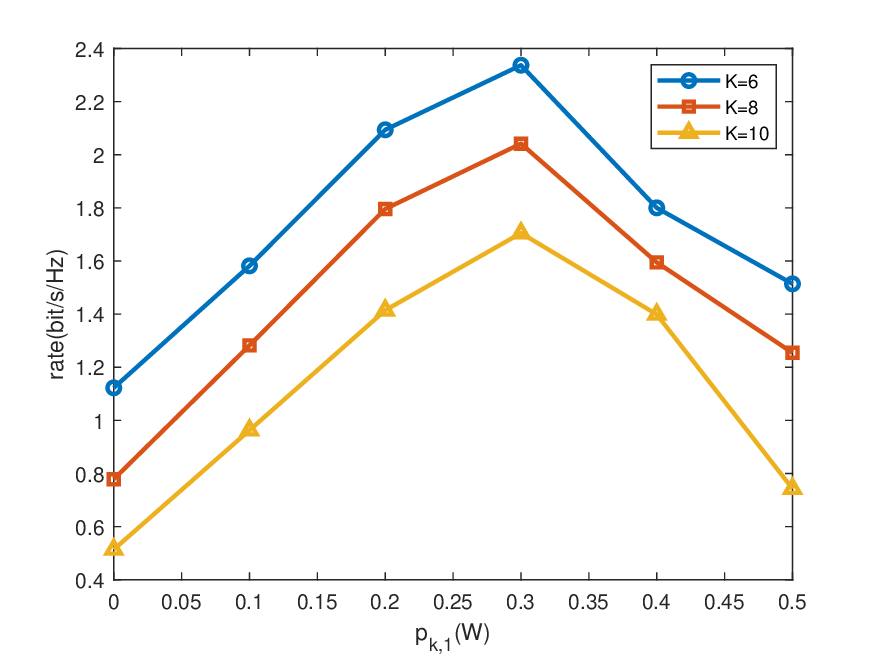}
	\end{center}
	\vspace{-0.4cm}
	\caption{The minimum sum rate for each device versus transmission power with different $K$. There is $L=4$, SNR$=10$dB, and $p_{k,1}+p_{k,2}=1$.}
	\label{fig2}
	\vspace{-0.5cm}
\end{figure}

Figure \ref{fig1} shows the minimum rate as the signal-to-noise ratio (SNR) increases with different $L$. It is observed that the minimum rate increases as SNR increases. In the meantime, increasing the number of groups $L$ makes the minimum rate higher. When $L=1$, the SGD is equal to the linear detection algorithm. It is evident that even if $L$ increases from $1$ to $2$, the minimum rate has been significantly increased. Therefore, SGD can strike a balance between linear detection and SIC in complexity and fairness. Figure \ref{fig2} illustrates the minimum sum rate for each device versus the transmit power. Given $p_{k,1}+p_{k,2}=1$, the numerical results show that the sum rate for each device is higher when $p_{k,1}/p_{k,2}=3/7$. It is worth noting that when $p_{k,1} = 0$, the system becomes a conventional NOMA. Moreover, the minimum rate decreases as the number of devices $K$ increases. It is because the interference is higher as $K$ becomes bigger. So it will be more important to apply SGD to improve the QoS of communication.

\section{Conclusion}

In this work, we study resource allocation in an uplink IRS-aided RSMA system. We apply SGD at the receiving end and construct an optimization problem. By optimizing the receive beamforming and phase-shift beamforming, the system can improve rate fairness for uplink communication. At the same time, the application of SGD also can provide reliable QoS. Therefore, the proposed scheme is valuable for ultra-reliability communication and is worth paying attention to it.

\appendices
\section{Proof of Theorem 1}
\label{appendix:t1}
Define the the objective function as $f(\{\mathbf G_{k,i}\}_{(k,i)\in\mathcal Q})$, i.e.,
\begin{eqnarray}
\begin{aligned}
f(\{\mathbf G_{k,i}\}_{(k,i)\in\mathcal Q})=r=\min\limits_{(k,i)\in \mathcal Q} u_{k,i}(\mathbf G_{k,i})-d_{k,i}^r(\mathbf G_{k,i}).\nonumber
\end{aligned}
\end{eqnarray}  
Assume that $\{\mathbf G_{k,i}^*\}_{(k,i)\in\mathcal Q}$ is the optimal solution of (P2). Then we have $f(\{\mathbf G_{k,i}\}_{(k,i)\in\mathcal Q})\le f(\{\mathbf G_{k,i}^*\}_{(k,i)\in\mathcal Q})$ for all $\mathbf G_{k,i}$ which satisfy $\text{tr}(\mathbf G_{k,i})-\|\mathbf G_{k,i}\|_2=0, \forall (k,i)\in \mathcal Q$. Let $g(\{\mathbf G_{k,i}\}_{(k,i)\in\mathcal Q},\rho_s)$ and $\{\mathbf G_{k,i}^s\}_{(k,i)\in\mathcal Q}$ denote the objective function and the optimal solution of (P2.1), respectively. With penalty factor $\rho_s$, there is 
\begin{eqnarray}
\begin{aligned}
g(\{\mathbf G_{k,i}^s\}_{(k,i)\in\mathcal Q},\rho_s)\ge g(\{\mathbf G_{k,i}^*\}_{(k,i)\in\mathcal Q},\rho_s),
\end{aligned}
\end{eqnarray}
which implies (\ref{eq:28}).
\begin{figure*}[!t]
	\vspace*{-8pt}
	\normalsize
	\begin{eqnarray}
	\begin{aligned}
	f(\{\mathbf G_{k,i}^s\}_{(k,i)\in\mathcal Q})-\frac{1}{2\rho_s}\left(\sum\limits_{(k,i)\in \mathcal Q}\text{tr}(\mathbf G_{k,i}^s)-\|\mathbf G_{k,i}^s\|_2\right)\ge f(\{\mathbf G_{k,i}^*\}_{(k,i)\in\mathcal Q})-\frac{1}{2\rho_s}\left(\sum\limits_{(k,i)\in \mathcal Q}\text{tr}(\mathbf G_{k,i}^*)-\|\mathbf G_{k,i}^*\|_2\right).
	\end{aligned} \label{eq:28}
	\end{eqnarray}
	\vspace*{-8pt}
\end{figure*}
Since $\{\mathbf G_{k,i}^*\}_{(k,i)\in\mathcal Q}$ is the optimal solution of (P2) and therefore the rank-one constraint must be satisfied, $\text{tr}(\mathbf G_{k,i}^*)-\|\mathbf G_{k,i}^*\|_2=0, \forall (k,i)\in \mathcal Q$. The above inequality is written as (\ref{eq:ap2}).
\begin{figure*}[!t]
	\vspace*{-10pt}
	\normalsize
	\begin{align}
	&f(\{\mathbf G_{k,i}^s\}_{(k,i)\in\mathcal Q})-\frac{1}{2\rho_s}\left(\sum\limits_{(k,i)\in \mathcal Q}\text{tr}(\mathbf G_{k,i}^s)-\|\mathbf G_{k,i}^s\|_2\right)\ge f(\{\mathbf G_{k,i}^*\}_{(k,i)\in\mathcal Q})\label{eq:ap2}\\&\Rightarrow\sum\limits_{(k,i)\in \mathcal Q}\text{tr}(\mathbf G_{k,i}^s)-\|\mathbf G_{k,i}^s\|_2\le 2\rho_s\left(f(\{\mathbf G_{k,i}^*\}_{(k,i)\in\mathcal Q})-f(\{\mathbf G_{k,i}^s\}_{(k,i)\in\mathcal Q})\right).\label{eq:ap1}
	\end{align}
	\vspace*{-10pt}
\end{figure*}
For $(k,i)\in \mathcal Q$, suppose $\bar{\mathbf G}_{k,i}$ is a limit point of sequence $\{{\mathbf G}_{k,i}^s\}$ and exist an infinite subsequence $\mathcal S$ such that $\lim_{s\in \mathcal S} \mathbf G_{k,i}^s = \bar{\mathbf G}_{k,i}$. By taking the limit as $s\to \infty, s\in \mathcal S$ on both side of (\ref{eq:ap1}), (\ref{eq:ap3}) holds,
\begin{figure*}[!t]
	\vspace*{-12pt}
	\normalsize
	\begin{eqnarray} 
	\begin{aligned}
	\sum\limits_{(k,i)\in \mathcal Q}\text{tr}(\bar{\mathbf G}_{k,i})-\|\bar{\mathbf G}_{k,i}\|_2&=\lim_{s\in \mathcal S}\left(\sum\limits_{(k,i)\in \mathcal Q}\text{tr}(\mathbf G_{k,i}^s)-\|\mathbf G_{k,i}^s\|_2\right)\\&\le \lim_{s\in \mathcal S}2\rho_s\left(f(\{\mathbf G_{k,i}^*\}_{(k,i)\in\mathcal Q})-f(\{\mathbf G_{k,i}^s\}_{(k,i)\in\mathcal Q})\right)\stackrel{\rho_s\to 0}{=}0.\label{eq:ap3}
	\end{aligned}
	\end{eqnarray}
	\hrulefill
	\vspace*{-8pt}
\end{figure*}
where the left side holds due to the continuity of function $\sum\limits_{(k,i)\in \mathcal Q}\text{tr}({\mathbf G}_{k,i})-\|{\mathbf G}_{k,i}\|_2$. In a result, there is $\sum\limits_{(k,i)\in \mathcal Q}\text{tr}(\bar{\mathbf G}_{k,i})-\|\bar{\mathbf G}_{k,i}\|_2=0$. So $\bar{\mathbf G}_{k,i}$ is feasible for (P2). By taking the limit as $s\to \infty, s\in \mathcal S$ on (\ref{eq:ap2}), we have 
\begin{align}
&f(\{\bar{\mathbf G}_{k,i}\}_{(k,i)\in\mathcal Q})\ge f(\{\bar{\mathbf G}_{k,i}\}_{(k,i)\in\mathcal Q})\nonumber\\&-\lim_{s\in \mathcal S}\frac{1}{2\rho_s}\left(\sum\limits_{(k,i)\in \mathcal Q}\text{tr}(\mathbf G_{k,i}^s)-\|\mathbf G_{k,i}^s\|_2\right)\ge f(\{\mathbf G_{k,i}^*\}_{(k,i)\in\mathcal Q})\nonumber
\end{align}
where $\rho_s$ and $\text{tr}(\mathbf G_{k,i}^s)-\|\mathbf G_{k,i}^s\|_2$ are non-negative.
Therefore, $\{\bar{\mathbf G}_{k,i}\}_{(k,i)\in\mathcal Q}$ is a set of feasible points whose objective value is no less than that of the optimal solution $\{\mathbf G_{k,i}^*\}_{(k,i)\in\mathcal Q}$. Therefore, $\{\bar{\mathbf G}_{k,i}\}_{(k,i)\in\mathcal Q}$ is also an optimal solution for (P2).

\ifCLASSOPTIONcaptionsoff
  \newpage
\fi

\section*{Acknowledgement}

This work is supported by National key project 2020YFB1807700, NSFC 62071296, Shanghai 22JC1404000, and PKX2021-D02.

\bibliographystyle{IEEEtran}
\bibliography{main}


\end{document}